# Magnetic Counting Rule of Radical Carbon Edge Nano Graphene

Norio Ota, Narjes Gorjizadeh* and Yoshiyuki Kawazoe*

Pure and Applied Sciences, University of Tsukuba, 1-1-1 Tenoudai, Tsukuba 305-8573 JAPAN

* Institute for Materials Research, Tohoku University, 2-1-1 Katahira, Aoba-ku, Sendai 980-8577 JAPAN

In order to explain room-temperature ferromagnetism of graphite-like materials, this paper offers a new magnetic counting rule of radical carbon zigzag edge nano graphene. Multiple spin state analysis based on a density function theory shows that the highest spin state is most stable. Energy difference with next spin state overcomes kT=2000K suggesting a room-temperature ferromagnetism. Local spin density at a radical carbon shows twice a large up-spin cloud which comes from two orbital with tetrahedral configuration occupied by up-up spins. This leads a new magnetic counting rule to give a localized spin Sz=+2/2 to one radical carbon site, whereas Sz= -1/2 to the nearest carbon site. Applied to five model molecules, we could confirm this magnetic counting rule. In addition, we enhanced such concept to oxygen substituted zigzag edge occupied by four electrons.

## 1. Introduction

Carbon based metal-free ferromagnetic materials are very attractive both in scientific aspect of π-electron oriented magnetism and both in technological aspect of light weight ecological magnet[1)-4)] and novel spintronic devices[5)-9)].

These years, several experiments[10)-15)] show room-temperature ferromagnetism in graphite like materials. Very recently, H.Ohldag et al.[15)] opened that proton ion implanted pure graphite shows a strong surface magnetism. Saturation magnetization is estimated to be 12-15emu/g at 300K by SQUID and XMCD measurements. This value is almost a quarter of metal nickel magnetization. Also T.Saito et al.[12)] developed ferromagnetic graphite like carbon material by a pyrolysis method from polyacrylonitrile, which has a saturation magnetization of 1.22emu/g at a room-temperature. Those experiments stimulate us to create new attractive materials as like an ultra light permanent magnet and a nano meter scale spintronic device. However, suitable models and guiding principles how to design those materials are not clear yet.

There were many theoretical predictions on graphene zigzag edge carbon magnetism due to localized density states near Fermi energy[16)-20)]. Focusing on infinite length graphene ribbon, Fujita et al[16)] applied tight binding model resulting antiferromagnetic feature with total magnetization zero. Rectangular shaped graphene has been extensively studied[21)-24)]. The singlet state (zero magnetization) arises by both side zigzag edge modifications with various species like -H, -F, -O,-OH, and -CH$_3$.[23)] Whereas, Kusakabe and Maruyama[25)26)] proposed an asymmetric infinite length ribbon model showing ferrimagnetic behavior with non-zero total magnetization.

Our previous papers[27)28)] reported multiple spin state analysis in asymmetric graphene molecules using a first principle density function theory (DFT). Especially, in a dihydrogenated zigzag edge graphene molecule, the most stable energy state was the highest spin state[28)]. Energy difference with next spin state overcomes kT=3000K.

In case of radical carbon zigzag edge molecule, such a detailed magnetic analysis is not applied yet. This paper opens multiple spin state behavior of radical carbon edge molecule. It should be noted that through such DFT analysis we could found a new magnetic counting rule. In polycyclic aromatic hydro carbon molecule (PAH), many chemists apply Lieb's rule[29)]. Maruyama et al[30)] proposed a rule to dihydrogenated zigzag edge graphene. However, those rules are not suitable to apply to radical carbon edge case. Here, we offer a new magnetic counting rule based both on DFT analysis and on molecular Hund's rule[31)32)]. In addition, magnetism of oxygen substituted zigzag edge molecule is well understood under a similar consideration.

## 2, Model Molecules

Proton ion irradiation experiment[10)] inspired us to make a proper model like piled graphene molecules as shown in Fig.1. Protons are irradiated from top. One side zigzag edge is modified to mono hydrogenated carbon (CH), whereas opposite side remains as radical carbon (C·) by a shadow effect.

Reported experiments mostly use powder samples, which suggest us modeling nano meter size molecule. Model molecules have neutral total charge, whereas have unpaired electrons to give magnetic polarization. Actual calculation models are five molecules illustrated in Fig.2 : (a) $C_{64}H_{17}$ with five radical carbons, (b) $C_{56}H_{16}$(four radical carbons), (c) $C_{64}H_{19}$(three), (d) $C_{56}H_{18}$ (two) and (e) $C_{64}H_{21}$(one) . In those molecules, there may appear complex multiple spin state. If the highest spin state had the energetically lowest stable energy and energy difference with next spin state overcome kT=1000K, we can believe room-temperature ferromagnetism

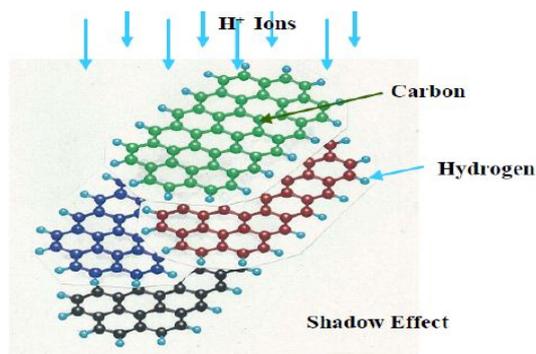

**Fig.1** Graphene molecules are piled one by one, where proton ions are irradiated from top. One side of edge carbon is mono-hydrogenated by proton irradiation, the other side remains radical carbon by a shadow effect.

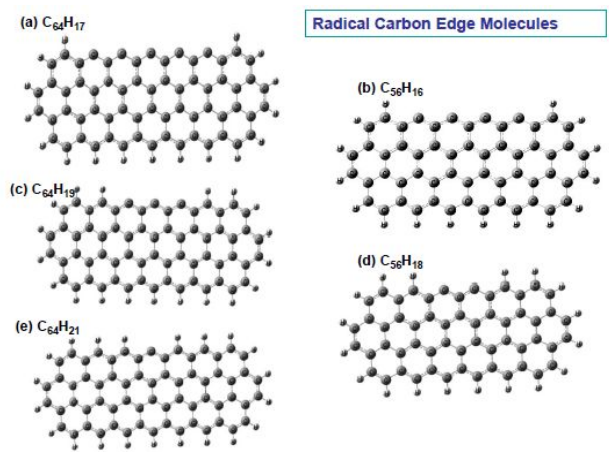

**Fig.2** Typical calculation model molecules with radical carbon zigzag edges: (a)$C_{64}H_{17}$, (b)$C_{56}H_{16}$, (c)$C_{64}H_{19}$, (d)$C_{56}H_{18}$ and (e)$C_{64}H_{21}$, where large ball show carbon atom and small one hydrogen.

### 3. Calculation Methods

In order to clarify magnetism, we have to obtain (i)spin density map, (ii) total molecular energy and (iii) optimized atom arrangement depending on respective given spin state Sz. Density function theory (DFT) [33)34)] based generalized gradient approximation method (GGA-UPBEPBE)[35)] is applied for those calculation. Atomic orbital basis is 6-31G [36)]. Energy accuracy is required at least 10E-8 as a total molecular energy after repeating atom position optimization.

### 4, Radical carbon edge molecule

Every molecule has several multiple spin state capabilities as like Sz=1/2, 3/2 and 5/2 in $C_{64}H_{17}$, Sz=0/2, 2/2 and 4/2 in $C_{56}H_{16}$ and so on. Those values are installed spin parameter as a given Sz to start DFT calculation. DFT Spin density, S(S+1) value and total molecular energy are calculated for every given Sz respectively. Results are summarized in Table 1. Given Sz(Sz+1) is a simple estimation of molecular total S square value S(S+1).Typical spin density maps are shown in Fig.3 comparing Sz=2/2 and 4/2 in $C_{56}H_{16}$, also Sz=1/2 and 3/2 in $C_{64}H_{19}$, where up-spins are shown in red(dark gray), down-spins in blue(light gray) at 0.001e/A$^3$ contour surface. Also in Fig.4, spin density maps of every highest spin state of five molecules are illustrated. Looking at those figures, we can notice following characteristics,

(1) In every molecule, the highest spin state shows regularly aligned spins, that is, up- and down-spins alternatively align one by one.
(2) Whereas, in case of lower spin state, there appears very complex spin structure inside of a molecule as like up-up spin pairs and down-down one.
(3) At the highest spin state, radical carbon exhibit very large up-spin local density, almost twice a large cloud.

**Table 1** DFT calculation results and estimated molecular spin Sz obtained by a new magnetic counting rule. Given Sz is an installed spin parameter to start DFT calculation. Given Sz(Sz+1) is a simple estimation of molecular S(S+1). ΔE is the energy difference from the lowest energy state.

| Chemical Formula | Given Sz | Given Sz(Sz+1) | DFT S(S+1) | DFT ΔE (*1) | Sz by a magnetic counting rule |
|---|---|---|---|---|---|
| $C_{64}H_{17}$ | 5/2 | 8.75 | 9.78 | 0.0 | 5/2 |
|  | 3/2 | 3.75 | 5.59 | 2.5 | — |
|  | 1/2 | 0.75 | 3.20 | 5.8 | — |
| $C_{56}H_{16}$ | 4/2 | 6.00 | 6.75 | 0.0 | 4/2 |
|  | 2/2 | 2.00 | 3.32 | 1.9 | — |
|  | 0/2 | 0.00 | (*2) | (*2) | — |
| $C_{64}H_{19}$ | 3/2 | 3.75 | 4.68 | 0.0 | 3/2 |
|  | 1/2 | 0.75 | 2.39 | 2.9 | — |
| $C_{56}H_{18}$ | 2/2 | 2.00 | 3.32 | 0.0 | 2/2 |
|  | 0/2 | 0.00 | (*2) | (*2) | — |
| $C_{64}H_{21}$ | 1/2 | 0.75 | 0.86 | 0.0 | 1/2 |

(*1) Energy difference from the lowest state(kcal/mol)
(*2) SCF calculation does not converged

Exchange coupling between up-up spins (also down-down) make the local exchange energy increase and finally elevates total molecular energy. Therefore, in every molecule, most stable spin state is assumed to be the highest one.

Exact DFT calculation was done. Obtained results are noted in Table 1 and Fig.5, where ΔE is the energy difference from the lowest energy state. For example, in

$C_{64}H_{27}$ among three spin states the lowest energy state is the highest spin one as Sz=5/2. Energy difference between Sz=5/2 and Sz=3/2 is 6.1kcal/mol. This value overcomes thermal excitation energy kT=2000K suggesting a room-temperature ferromagnetism

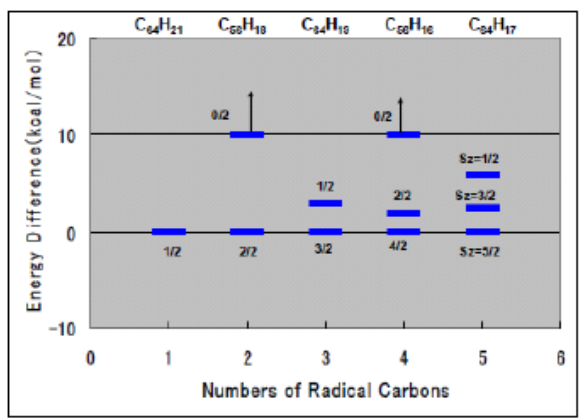

**Fig.5** Energy difference between spin states for radical carbon edge molecule. In every molecule, the highest spin state is energetically most stable.

### 5, Magnetic Counting Rule of Radical Carbon

There is a famous magnetic counting rule on polycyclic aromatic hydrocarbon (PAH) named Lieb's rule[28]. One example is shown in Fig.6. Carbons are classified to two groups, one is A site with up-spin (Sz=+1/2), another is B site with down-spin (Sz= -1/2). Those A and B sites are alternately aligned one by one. In case of simple model like $C_{14}H_{10}$, there are seven A sites and seven B sites. Total spin is canceled each other to become Sz=0, which means diamagnetic nature of conventional PAH.

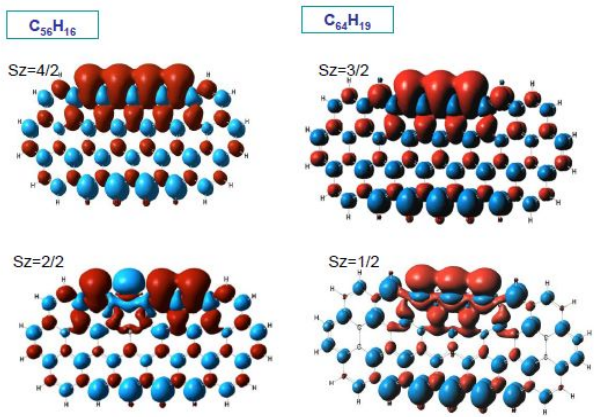

**Fig.3** Spin density map of Sz=2/2 and 4/2 in $C_{56}H_{16}$, also Sz=1/2 and 3/2 in $C_{64}H_{19}$, where up spins are shown in red(dark gray), down spins in blue(light gray) at $0.001e/A^3$ contour surface.

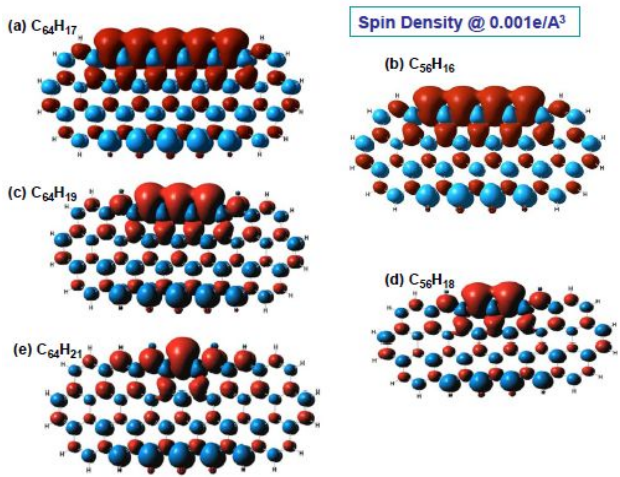

**Fig.4** Spin density map of every highest spin state for five molecules. In every molecule, edge radical carbon has almost twice a large up spin density.

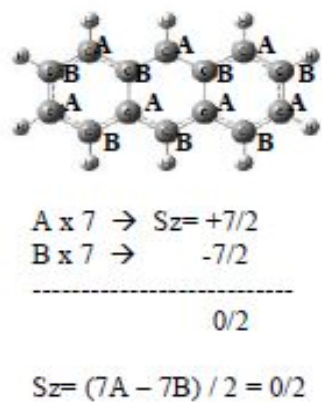

**Fig.6** Example of magnetic counting rule on polycyclic aromatic hydrocarbon (PAH). Following to Lieb's rule[29], carbons are classified to two groups as A or B having Sz=+1/2 or -1/2 spin respectively. Total molecular spin is a simple subtraction of A and B numbers.

In case of radical carbon edge molecule, we cannot apply such simple counting rule. Example is shown in Fig.7 (a) as $C_{13}H_7$ small molecule with two zigzag edge radical carbons. If apply Lieb's rule, stable Sz may be 1/2. However, DFT calculation result is Sz= 3/2. We should propose a new rule for radical carbon case. Spin density distribution is illustrated in (b) as a plane view and in (c) as a side view. Almost twice a large up spin density is mapped at a radical carbon site. As imaged in (d), we can suppose two orbits (triangle marks) with tetrahedral configuration. If two identical electrons occupy those orbits, there appears up-up parallel spin pair under the Hund's rule [31)32)]. Therefore, we can propose a new rule, that is, one radical carbon site should be assigned to be 2A (Sz=2/2). In case of $C_{13}H_7$, there are two 2A, five A and six B. Total molecular spin is counted as a subtraction of those A and B numbers resulting (9A-6B) to be Sz=+3/2, which is consistent with detailed DFT calculation.

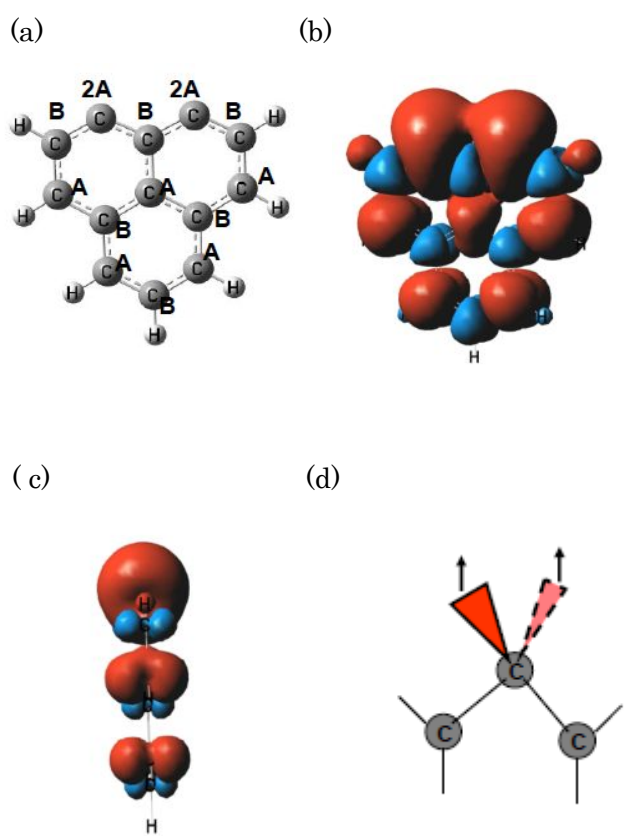

Fig.7 New magnetic counting rule applied to radical carbon edge molecule: (a) $C_{13}H_7$ molecule example labeled 2A for radical carbon site, (b) spin density plane view, (c) side view, (d) schematic orbital image with tetrahedral configuration. By the Hund's rule, two electrons occupy those orbits as up-up spin pair.

We applied this new magnetic counting rule to a larger molecule. An example is shown in Fig.8 for $C_{64}H_{17}$ with five radical carbons. There are five 2A, twenty seven A and thirty two B. Total molecular Sz is a simple subtraction of total A and B resulting 5A(Sz=5/2). Also we applied this rule to other molecules. Estimated Sz by a magnetic counting rule are summarized in Table 1. All of counted Sz completely agree with DFT calculated most stable Sz.

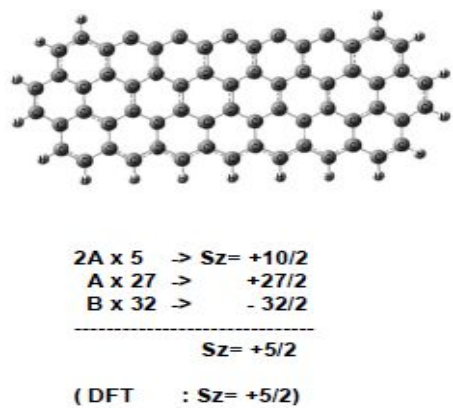

Fig.8 Total molecular spin Sz of $C_{64}H_{17}$ applying a new magnetic counting rule. Radical carbon edge is assigned to be 2A. Total molecular spin is a subtraction of A and B numbers resulting Sz=+5/2, which is consistent with DFT calculation.

### 6, Oxygen Substituted ZigZag Edge

Tetrahedral configuration of zigzag edge site electrons lead us a question, that is, what happen in case of four electrons. Simple answer is illustrated in Fig.9. Oxygen substituted zigzag edge realizes such situation. Two electrons occupy one orbital as up-down spin pair by Pauli principle. Additional two electrons occupy identical another orbit also as up-down pair. Total spin is cancelled each other. Stable Sz is zero at oxygen site.

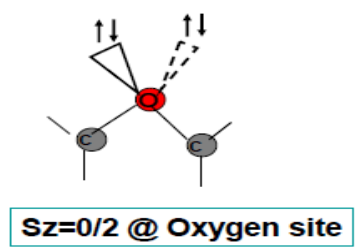

Fig.9 Four electrons molecular orbit with tetrahedral configuration. Up-down spin pair occupy one orbit. Identical another orbit is also occupied by extra two electrons. Total local spin Sz=0/2. Such situation will be realized by oxygen substitution.

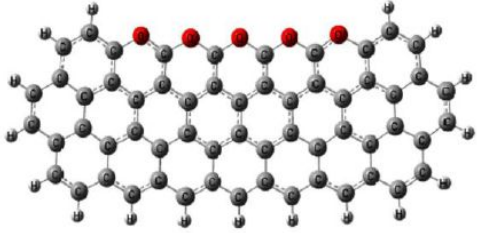

($C_{59}O_5H_{17}$ molecule: Oxygen atom is red (dark gray) ball, carbon large light gray and hydrogen small one)

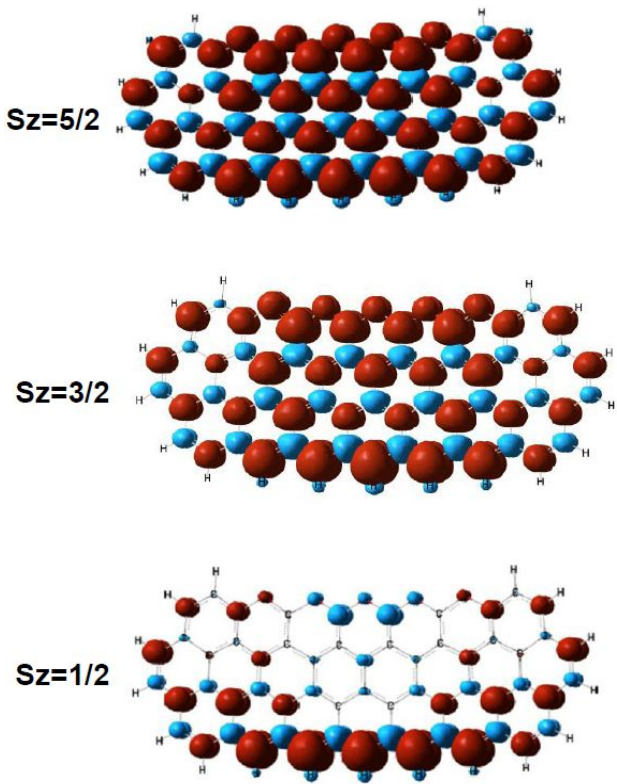

**Fig.10** Spin density map of $C_{59}O_5H_{17}$ molecule for Sz=5/2, 3/2 and 1/2. There are serious up-up spin pairs for higher spin state. Lowest spin state is energetically favorable and stable.

Such simple speculation is checked by DFT calculation. Fig.10 show spin density of $C_{59}O_5H_{17}$ for Sz=5/2, 3/2 and 1/2 cases respectively. There are many and strong up-up spin pairs in Sz=5/2 and 3/2. Those bring unfavorable energy increase. Detailed energy difference is calculated and illustrated in Fig.11. The lowest spin state is most stable. This result suggests a diamagnetic or weak magnetic nature of oxygen substituted graphene molecule.

.

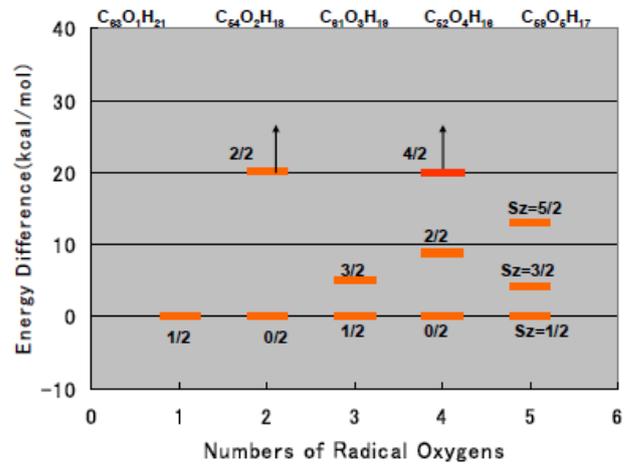

**Fig.11** Molecular energy comparison for five molecules with Oxygen substituted zigzag edges. In every molecule, the lowest spin state is energetically most stable.

### 5, Conclusion

Recent experiments indicate room-temperature ferromagnetism in graphite-like materials. In order to explain those experiments, this paper newly proposes a magnetic counting rule of radical carbon edge graphene molecule based on a first principle density function theory (DFT) analysis. Molecules with radical carbon zigzag edges like $C_{64}H_{17}$, $C_{56}H_{16}$, $C_{64}H_{19}$, $C_{56}H_{18}$ and $C_{64}H_{21}$ show that in all molecules the highest spin state is most stable with kT=2000K energy difference with next spin state. This result suggests a stability of room-temperature ferromagnetism. Spin density at a radical carbon edge shows twice a large up-spin cloud which originates from two unpaired electrons of radical carbon. Two electrons orbit show tetrahedral configuration with identical energy and obey to Hund's rule. Such quantum analysis leads a new magnetic counting rule giving 2A (Sz=+2/2) to one radical carbon site, B (Sz=-1/2) to the nearest carbon and simply add those numbers to estimate total molecular spin Sz. Appling this counting rule to above mentioned five molecules, we could reproduce all the stable spin states agree with DFT result. In addition, we applied four electrons case which means oxygen atom substitution to zigzag edge site. One orbital is occupied by up-down spin pair, another identical orbital is also occupied by additional up-down pair. By such simple estimation, oxygen modified molecule may exhibit stable spin state to be the lowest one. This is also confirmed by detailed DFT calculation on five molecules with different numbers of oxygen atom.

Acknowledgements

Narjes Gorjizadeh would like to thank the crew of the Center for Computational Materials Science, Institute for Materials Research of Tohoku University for their support of the Hitachi SR11000(model K2) supercomputer system, and Global COE Program "Materials Integration (International Center of Education and Research),Tohoku University," MEXT, Japan, for financial support.

References

1) A.A.Ovchinnikov and V.N.Specto:
Synth.Met., **27**, B615 (1988)
2) K.Murata, H.Ueda and K.Kawaguchi:
Synth.Met., **44**, 357 (1991)
3) K.Murata, H.Ushijima, H.Ueda and K.Kawaguchi : J. Chem. Soc.,Chem. Commun., **7**,567 (1992)
4) J.Miller: Inorg.Chem., **39**, 4392(2000)
5) Y.Shiroishi, F.Fukuda, I.Tagawa, H.Iwasaki, S.Takenoiri, H.Tanaka, M.Mutoh and N.Yoshikawa: IEEE Trans. On Magnetics, 45, 3816 (2009)
6) Y.Shiroishi : まぐね,**5,** (2010) ,Review in Japanese
7) M.Ohishi, M.Shiraishi, R.Nouchi, T.Nozaki, T.Shinjyo and Y.Suzuki : Jpn.J.Appl.Phys. **46**, L605 (2007)
8) 白石誠司: 表面科学、**29**、310（2008）,Review in Japanese
9) N.Tombros, C.Jozsa, M.Popinciuc, H.T.Jonkman and B.J.van Wees: Nature, **448**, 571 (2007)
10) P.Esquinazi, D.Spemann, R.Hohne, A.Setzer, K.Han, and T.Butz: *Phys.Rev.Lett.,* **91**, 227201(2003)
11) K.Kamishima, T.Noda, F.Kadonome, K.Kakizaki and N.Hiratsuka: ,J.of Magnetism and Magnetic Materials **310** , e346(2007)
12) T.Saito, D.Nishio-Hamane, S.Yoshii, and T.Nojima: Appl.Phys.Letters, **98**, 052506 (2011)
13) Y.Wang,Y.Huang,Y.Song,X.Zhang,Y.Ma,J.Liang and Y.Chen: Nano Letters ,**9**, 220(2009)
14) J.Cervenka, M.Katsnelson and C.Flips: Nature Physics online Oct.04 (2009)

15) H.Ohldag,P.Esquinazi,E.Arenholz,D.Spemann,M.Rothermal, A.Setzer, and T.Butz: New Journal of Physics, **12**, 123012 (2010)
16) M.Fujita,K.Wakabayashi, K.Nakada and K.Kusakabe: J.of the Phys.Soc.of Japan .**65**, 1920(1996)
17) K.Nakada,M.Fujita,G.Dresselhaus and M.Dresselhaus: Phy.Rev. B **54**, 17954(1996)
18) K.Wakabayashi,M.Fujita,H.Ajiki and M.Sigrist: Phys.Rev. B **59**, 8271(1999)
19) Y.Miyamoto,K.Nakada and M.Fujita: Phys.Rev.B, **59**, 9858(1999)
20) J.Berashevich and T. Chakraborty: Phys.Rev.B, **80,** 115430 (2009)
21) H.Zheng and W.Duley: Phys.Rev.B, **78**, 045421(2008)
22) D.E.Jiang,B.G.Sumpter and S.Dai: J.Chem.Phys., **127** , 124703(2007)
23) X.Gao,Z.Zhuo,Y.Zhao,S.Nagase,S.B.Zhang and Z.Chen: J.Phys.Chem.C**112**, 12677(2008)
24) T.Wassman,A.P.Seitsonen,A.M.Saitta,M.Lazzeri and F.Mauri: Phys.Rev.Lett., **101**, 096402(2008)
25) K.Kusakabe and M.Maruyama: Phys.Rev.B, **67**, 092406(2003)
26) M.Maruyama and K.Kusakabe: J.of the Phys.Soc.of Japan , .**73**, 656 (2004)
27) N.Ota, N.Gorjizadeh and Y.Kawazoe : Journal of the Magnetics Society of Japan **34,** 573 (2010)
28) N.Ota, N.Gorjizadeh and Y.Kawazoe :
Preprint is available at arXiv web site, **arXiv:**1101.4080
29) E.H.Lieb : Phys.Rev.Lett. 62,1201 (1989)
30) M.Maruyama, K.Kusakabe, S.Tsuneyuki, K.Akagi, Y.Yoshimoto and J.Yamauchi : J. of Physics and Chemistry of Solids **65**,119 (2004)
31) F.Hund : Z Physik **33,** 345 (1925)
32) Y.Maruyama, K.Hongo, M.Tachikawa, Y.Kawazoe and H.Yasuhara : Internatinal J. of Quantum Chemistry, **108**, 731 (2008)
33) P.Hohenberg and W.Kohn: Phys.Rev., **136** , B864(1964)
34) W.Kohn and L.Sham: Phys.Rev., **140**, A1133(1965)
35) J.P.Perdew, K.Burk and M.Ernzerhof: Phy.Rev.Lett., 77, 3865(1966)
36) R.Ditchfield, W.Hehre and J.Pople : J.Chem.Phys. **54**, 724(1971)